%% file: WZPaper.tex
\RequirePackage{lineno}

\pdfoutput=1
\documentclass[preprint,5p]{elsarticle}

\usepackage{subfigure}
\usepackage{atlasphysics}
\usepackage{booktabs} 
\usepackage{epstopdf}
\usepackage{lineno}
\usepackage{textpos}
\biboptions{sort&compress}
\usepackage{hyperref}

\input{ResultsDB}

\journal{Physics Letters B} 
\begin{document}

\begin{frontmatter}

\title{\vspace{-0.5cm} \flushright{\normalsize{CERN-PH-EP-2011-184}} \\ \center{Measurement of the \WZ Production Cross Section and Limits on Anomalous Triple Gauge Couplings in Proton-Proton Collisions at ${\sqrt{s} =7}\TeV$ with the ATLAS Detector}}
\author{The ATLAS Collaboration}

\begin{abstract}
This Letter presents a measurement of \WZ production in  \lumi~\ifb\ of $pp$ collision data
at $\sqrt s = 7\TeV$ collected by the ATLAS experiment in 2011. Doubly leptonic decay events are selected with electrons, muons and missing transverse momentum in the final state.
In total \WZNSigEventsObserved\ candidates are observed, with a background expectation of \WZNBkgEventsExpected$\pm$\WZNBkgEventsExpectedErrStat(stat.)$^{+\WZNBkgEventsExpectedErrUpSys}_{-\WZNBkgEventsExpectedErrDwSys}$(syst.) events. 
The total cross section for \WZ production for $\Z/\gamma^*$ masses within the range 66~GeV to 116~GeV is determined to be 
$\sigma_{W\!Z}^\mathrm{tot}=  \WZtotXsec^{+\WZtotXsecStatErrUp}_{-\WZtotXsecStatErrDw}$(stat.)$^{+\WZtotXsecSysErrUp}_{-\WZtotXsecSysErrDw}$(syst.)$^{+\WZtotXsecLumiErrUp}_{-\WZtotXsecLumiErrDw}$(lumi.)~\pb, which is consistent with the Standard Model expectation of $\WZtotXsecExpected^{+\WZtotXsecExpectedErrUp}_{\WZtotXsecExpectedErrDw}$~\pb. Limits on anomalous triple gauge boson couplings are extracted.
\end{abstract}

\end{frontmatter}

\section{Introduction}\label{sec:Introduction}
\input{Introduction}

\section{The ATLAS Detector and Event Samples}\label{sec:ATLAS}
\input{ATLASandDataSample}

\input{Trigger}

\input{MCSamples}
\section{Object Reconstruction}\label{sec:Reconstruction}
\input{ObjectReconstruction}
\section{Event Selection}\label{sec:EventSelection}
\input{EventSelection}
\section{Signal Efficiency and Background Estimate}\label{sec:SignalAcceptance}
\input{SignalAcceptance}

\input{Systematics}

\input{Background}

\section{Results}\label{sec:Results}
\input{Results}

\section{Acknowledgements}

We thank CERN for the very successful operation of the LHC, as well as the
support staff from our institutions without whom ATLAS could not be
operated efficiently.

We acknowledge the support of ANPCyT, Ar\-gen\-tina; YerPhI, Ar\-me\-nia; ARC,
Au\-stra\-lia; BMWF, Au\-stria; AN\-AS, Azer\-bai\-jan; SSTC, Be\-la\-rus; CNPq and FAPESP,
Bra\-zil; NSERC, NRC and CFI, Canada; CERN; CONICYT, Chi\-le; CAS, MOST and
NSFC, China; COLCIENCIAS, Colombia; MSMT CR, MPO CR and VSC CR, Czech
Republic; DNRF, DNSRC and Lundbeck Foundation, Denmark; ARTEMIS, European
Union; IN2P3\--CNRS, CEA-DSM/IRFU, France; GNAS, Georgia; BMBF, DFG, HGF, MPG
and AvH Foundation, Germany; GSRT, Greece; ISF, MINERVA, GIF, DIP and
Benoziyo Center, Israel; INFN, Italy; MEXT and JSPS, Japan; CNRST, Morocco;
FOM and NWO, Netherlands; RCN, Norway; MNiSW, Poland; GRICES and FCT,
Portugal; MERYS (MECTS), Romania; MES of Russia and ROSATOM, Russian
Federation; JINR; MSTD, Serbia; MSSR, Slovakia; ARRS and MVZT, Slovenia;
DST/NRF, South Africa; MICINN, Spain; SRC and Wallenberg Foundation,
Sweden; SER, SNSF and Cantons of Bern and Geneva, Switzerland; NSC, Taiwan;
TA\-EK, Turkey; STFC, the Royal Society and Leverhulme Trust, United Kingdom;
DOE and NSF, United States of America.

The crucial computing support from all WLCG partners is acknowledged
gratefully, in particular from CERN and the ATLAS Tier-1 facilities at
TRIUMF (Canada), NDGF (Denmark, Norway, Sweden), CC-IN2P3 (France),
KIT / GridKA (Germany), INFN-CNAF (Italy), NL-T1 (Netherlands), PIC (Spain),
ASGC (Taiwan), RAL (UK) and BNL (USA) and in the Tier-2 facilities
worldwide.

\bibliographystyle{apsrev4-1}
\bibliography{WZPaper}

\clearpage
\onecolumn
\input{atlas_authlist}

\clearpage

\end{document}

%% file: ResultsDB.tex
\def\lumi{1.02} 

\def\lumiAbsErr{0.04} 

\def\WZtotXsecExpected{17.3}
\def\WZtotXsecExpectedErrUp{1.3}
\def\WZtotXsecExpectedErrDw{-0.8}

\def\WZNSigEventsExpected{50.3}
\def\WZNSigEventsExpectedErrStat{0.4}

\def\WZNSigEventsExpectedErrSys{4.3}

\def\WZNSigEventsObserved{71}

\def\WZNSigEventsExpectedEEE{7.7}
\def\WZNSigEventsExpectedErrStatEEE{0.2}

\def\WZNSigEventsObservedEEE{11}

\def\WZNSigEventsExpectedEEM{11.6}
\def\WZNSigEventsExpectedErrStatEEM{0.2}

\def\WZNSigEventsObservedEEM{9}

\def\WZNSigEventsExpectedEMM{12.4}
\def\WZNSigEventsExpectedErrStatEMM{0.2}

\def\WZNSigEventsObservedEMM{22}

\def\WZNSigEventsExpectedMMM{18.6}
\def\WZNSigEventsExpectedErrStatMMM{0.3}

\def\WZNSigEventsObservedMMM{29}

\def\WZNBkgEventsExpected{12.1}

\def\WZNBkgEventsExpectedErrStat{1.4}

\def\WZNBkgEventsExpectedErrUpSys{4.1}
\def\WZNBkgEventsExpectedErrDwSys{2.0}
\def\WZNZZBkgEventsExpected{3.9}
\def\WZNZZBkgEventsExpectedErrStat{0.1}
\def\WZNZZBkgEventsExpectedErrSys{0.2}
\def\WZNTopBkgEventsExpected{2.3}
\def\WZNTopBkgEventsExpectedErrStat{1.0}
\def\WZNTopBkgEventsExpectedErrSys{0.5}
\def\WZNZgammaBkgEventsExpected{1.1}
\def\WZNZgammaBkgEventsExpectedErrStat{0.5}
\def\WZNZgammaBkgEventsExpectedErrSys{0.1}
\def\WZNZjetsDDBkgEventsExpected{4.8}
\def\WZNZjetsDDBkgEventsExpectedErrStat{0.8}
\def\WZNZjetsDDBkgEventsExpectedErrSysUp{4.0}
\def\WZNZjetsDDBkgEventsExpectedErrSysDown{1.9}

\def\WZNBkgEventsExpectedEEE{3.1}
\def\WZNBkgEventsExpectedErrStatEEE{0.6}

\def\WZNTopBkgEventsExpectedEEE{0.2}
\def\WZNTopBkgEventsExpectedErrStatEEE{0.1}

\def\WZNZZBkgEventsExpectedEEE{0.4}
\def\WZNZZBkgEventsExpectedErrStatEEE{0.0}

\def\WZNZgammaBkgEventsExpectedEEE{0.5}
\def\WZNZgammaBkgEventsExpectedErrStatEEE{0.3}

\def\WZNZjetsDDBkgEventsExpectedEEE{2.0}
\def\WZNZjetsDDBkgEventsExpectedErrStatEEE{0.5}

\def\WZNBkgEventsExpectedEEM{2.5}
\def\WZNBkgEventsExpectedErrStatEEM{0.7}

\def\WZNTopBkgEventsExpectedEEM{0.8}
\def\WZNTopBkgEventsExpectedErrStatEEM{0.6}

\def\WZNZZBkgEventsExpectedEEM{1.0}
\def\WZNZZBkgEventsExpectedErrStatEEM{0.1}

\def\WZNZjetsDDBkgEventsExpectedEEM{0.7}
\def\WZNZjetsDDBkgEventsExpectedErrStatEEM{0.3}

\def\WZNBkgEventsExpectedEMM{3.9}
\def\WZNBkgEventsExpectedErrStatEMM{0.9}

\def\WZNTopBkgEventsExpectedEMM{0.9}
\def\WZNTopBkgEventsExpectedErrStatEMM{0.7}

\def\WZNZZBkgEventsExpectedEMM{0.8}
\def\WZNZZBkgEventsExpectedErrStatEMM{0.1}

\def\WZNZgammaBkgEventsExpectedEMM{0.6}
\def\WZNZgammaBkgEventsExpectedErrStatEMM{0.4}

\def\WZNZjetsDDBkgEventsExpectedEMM{1.7}
\def\WZNZjetsDDBkgEventsExpectedErrStatEMM{0.5}

\def\WZNBkgEventsExpectedMMM{2.6}
\def\WZNBkgEventsExpectedErrStatMMM{0.6}

\def\WZNTopBkgEventsExpectedMMM{0.4}
\def\WZNTopBkgEventsExpectedErrStatMMM{0.5}

\def\WZNZZBkgEventsExpectedMMM{1.7}
\def\WZNZZBkgEventsExpectedErrStatMMM{0.1}

\def\WZNZjetsDDBkgEventsExpectedMMM{0.4}
\def\WZNZjetsDDBkgEventsExpectedErrStatMMM{0.3}

\def\AWZ{0.342}
\def\AWZErrorComb{0.006} 
\def\AWZrelErrorComb{1.5} 
\def\AWZrelErrorStat{0.6} 

\def\CWZeeeRel{34.3}
\def\CWZeeeRelErr{0.8} 

\def\CWZeemRel{50.2}
\def\CWZeemRelErr{0.9} 

\def\CWZemmRel{54.5}
\def\CWZemmRelErr{1.0} 

\def\CWZmmmRel{81.6}
\def\CWZmmmRelErr{1.3} 

\def\expglow{-0.12}
\def\expgup{0.20}
\def\expkappalow{-0.6}
\def\expkappaup{0.8}
\def\explambdalow{-0.11}
\def\explambdaup{0.11}

\def\obsglow{-0.16} 
\def\obsgup{0.24}   
\def\obskappalow{-0.8} 
\def\obskappaup{1.0}   
\def\obslambdalow{-0.14} 
\def\obslambdaup{0.14}   

\def\obsglowtwotev{-0.20} 
\def\obsguptwotev{0.30}   
\def\obskappalowtwotev{-0.9} 
\def\obskappauptwotev{1.1}   
\def\obslambdalowtwotev{-0.17} 
\def\obslambdauptwotev{0.17}   

\def\WZfidXsec{102} 
\def\WZfidXsecStatErrUp{15} 
\def\WZfidXsecStatErrDw{14} 
\def\WZfidXsecSysErrUp{7} 
\def\WZfidXsecSysErrDw{6} 
\def\WZfidXsecLumiErrUp{4} 
\def\WZfidXsecLumiErrDw{4}

\def\WZtotXsec{20.5} 
\def\WZtotXsecStatErrUp{3.1} 
\def\WZtotXsecStatErrDw{2.8} 
\def\WZtotXsecSysErrUp{1.4} 
\def\WZtotXsecSysErrDw{1.3} 
\def\WZtotXsecLumiErrUp{0.9} 
\def\WZtotXsecLumiErrDw{0.8}

\def\WZNSumSigEventsExpectedEEE{10.9}
\def\WZNSumSigEventsExpectedErrStatEEE{0.6}    
\def\WZNSumSigEventsExpectedEEM{14.0}
\def\WZNSumSigEventsExpectedErrStatEEM{0.7} 
\def\WZNSumSigEventsExpectedEMM{16.4}
\def\WZNSumSigEventsExpectedErrStatEMM{1.0} 
\def\WZNSumSigEventsExpectedMMM{21.2}
\def\WZNSumSigEventsExpectedErrStatMMM{0.7}
\def\WZSumNSigEventsExpected{62.4}
\def\WZNSumSigEventsExpectedErrStat{1.5}
\def\WZNSigEventsExpectedErrSysUp{5.9}
\def\WZNSigEventsExpectedErrSysDw{4.6}

%% file: Introduction.tex
The underlying structure of the electroweak interactions in the Standard Model (SM) is the non-abelian $SU(2)_{L}$ $\times$ $U(1)_{Y}$ gauge group. 
Properties of electroweak gauge bosons such as their masses and couplings to fermions have been precisely measured at LEP and the Tevatron~\cite{Precision}. However, triple gauge boson couplings (TGC) predicted by this theory have not yet been determined with comparable precision. 

In the SM the triple gauge boson vertex is completely fixed by the electroweak gauge structure. A measurement of this vertex, for example through the analysis of diboson production at the LHC, tests the gauge symmetry
and probes for possible new phenomena involving gauge bosons. 
In general, electroweak boson couplings deviating from gauge constraints yield enhancements of the \WZ production cross section at high diboson invariant mass.
Furthermore, new particles decaying into \WZ pairs are predicted in models with extra vector bosons (e.g. $W^{\prime}$) as well as in supersymmetric models with an extended Higgs sector (charged Higgs)~\cite{DKS:1999, Higgs}.

At the LHC, the dominant \WZ production mechanism is from quark-antiquark and quark-gluon interactions at leading order (LO) and at next-to-leading order (NLO), 
respectively~\cite{PhysRevD.65.094041}. 
Only the $s$-channel diagram has a triple electroweak gauge boson interaction vertex and is hence the only channel that may contribute to anomalous TGC (aTGC).

This Letter presents a measurement of the \WZ production cross section and limits on aTGC with the ATLAS detector in LHC proton-proton collisions at a centre-of-mass energy, $\sqrt{s}$, of 7 TeV. 
The analysis uses four channels with leptonic decays ($\WZ\to\ell\nu\ell\ell$) involving electrons and muons: $e\nu ee$, $\mu\nu ee$, $e\nu\mu\mu$ or $\mu\nu\mu\mu$, where the $\nu$ is estimated 
by the missing transverse momentum, $\MET$.
The main sources of background are $ZZ$, $Z\gamma$, $Z$+jets, and top-quark events. 

A common phase space is defined for combining the four decay
channels and measuring a ``fiducial'' cross section. 
The phase space is chosen to match closely the
detector acceptance and analysis selection. 
The leptons from the $Z$ and $W$ boson decays are required to have 
transverse momenta 
$\pt^{\mu,e}(\Z) > 15\GeV$,
$\pt^{\mu,e}(\W) > 20\GeV$,
pseudorapidity\footnote{ATLAS uses a right-handed coordinate
system with its origin at the nominal interaction point in the centre of the detector and
the $z$-axis along the beam pipe. The $x$-axis points from the interaction point to
the centre of the LHC ring, and the $y$-axis points upwards. Cylindrical coordinates ($r$,$\phi$)
are used in the transverse plane, $\phi$ being the azimuthal angle around the beam pipe.
The pseudorapidity $\eta$ is defined in terms of the polar angle $\theta$ as
$\eta = - \ln\tan(\theta/2)$.} 
$|\eta^{\mu,e}| < 2.5$, $|m_{\ell\ell}(Z)-m_Z| < 10\GeV$, $\pt^{\nu} > 25\GeV$
and the transverse mass\footnote{The transverse mass is defined as $\MT^2 = 2E_{\mathrm T}^{\ell}E_{\mathrm T}^{\nu}-2{\bf p}_{\mathrm T}^{\ell}{\bf p}_{\mathrm T}^{\nu}$.} 
of the $W$ boson is required to satisfy $m_{\mathrm{T}}^{W} > 20\GeV$. Final state electrons and muons whose four-momenta include all photons within $\Delta R<0.1$ are used in the phase space definition\footnote{$\Delta R$ is defined as $\Delta R = \sqrt{(\Delta \eta)^2 + (\Delta \phi)^2}$.}.
Since the fiducial phase space is defined by the lepton kinematics, the
cross section definition includes the branching ratios of the bosons
decaying into electrons or muons.  The fiducial cross section definition excludes the contribution from $W$ and $Z$ boson decays into $\tau$ leptons. 

In order to measure the total cross section, the experimentally accessible phase space is extrapolated
to the full phase space. The region dominated by the contribution of
a $\gamma^{*}$ propagator in singly resonant diagrams to the theoretical cross section is highly suppressed by requiring the invariant mass of the dilepton
system from $Z/\gamma^{*}$ to satisfy $66\GeV < m_{\ell\ell} < 116\GeV$ for the full phase space.

In the SM the only allowed boson combinations for TGC vertices are $WW\gamma$ and $WWZ$, and the latter is addressed in this Letter.
Expressions for the most general effective Lagrangian for a TGC vertex with two charged and one neutral vector boson can be found in Refs. \cite{Hagiwara} and \cite{EllisonWudka}. 
If only terms that separately conserve charge conjugation and parity are considered, then the couplings can be represented by three dimensionless parameters  $g^{Z}_{1}$, $\kappa_{Z}$ and $\lambda_{Z}$.  In the SM $g_{1}^{Z}=1$, $\kappa_{Z}=1$ and $\lambda_{Z}=0$. Anomalous couplings, defined as deviations from these SM values, are then $\Delta g^{Z}_{1}$, $\Delta\kappa_{Z}$ and $\lambda_{Z}$.

To avoid tree-level unitarity violation, which occurs in the effective Lagrangian approach at sufficiently large energies, 
the anomalous couplings must be suppressed 
at higher energy scales. 
To achieve this, an arbitrary form factor can be introduced to mitigate the effect of anomalous couplings at higher energy scales.
For comparison with previous studies, results are presented using 
a dipole form factor $f(\hat s)=1/(1+\hat s /\Lambda^2)^2$, where  $\Lambda=2\TeV$ is a cut-off energy scale and $\sqrt {\hat s}$ is the partonic 
centre-of-mass energy. This choice ensures that unitarity is not violated.
However, since the choice of the scale is arbitrary and the experimental centre-of-mass energy scale is finite, 
the interpretation of the data in the framework of anomalous couplings is also presented without using a form factor,
corresponding to setting $\Lambda=\infty$.

%% file: ATLASandDataSample.tex
The ATLAS detector~\cite{atlas} consists of an inner detector (ID)
surrounded by a superconducting solenoid which provides a 2\,T magnetic field, electromagnetic and
hadronic calorimeters and a muon spectrometer (MS) with a toroidal magnetic field. The ID
provides precision charged particle tracking for $|\eta|<2.5$. It consists of a
silicon pixel detector, a silicon strip detector and a straw tube tracker that also provides
transition radiation measurements for electron identification. The calorimeter system  covers the range $|\eta| < 4.9$
and comprises sampling calorimeters with either liquid argon (LAr) or scintillating tiles as the active media.
In the region $|\eta|<2.5$ the electromagnetic LAr
calorimeter is finely segmented and plays an important role in electron identification.
The muon spectrometer has separate trigger and high-precision tracking chambers which provide muon
identification in $|\eta|<2.7$.

This study uses $\lumi\pm\lumiAbsErr~\ifb$~\cite{bib:ATLASlumi,ATLAS-lumi2011} of collision data collected up to the end of June 2011.

%% file: Trigger.tex
Candidate events are selected online with single-lepton 
triggers requiring \pT\ of at least 18 (20) GeV for muons (electrons). 
The trigger efficiency for $\WZ\rarrow \ell\nu\ell\ell $  events
which pass all selection criteria is in the range of 96--99\%
depending on the final state.

%% file: MCSamples.tex
The \WZ production processes and the subsequent purely leptonic decays
are modelled by the \mcatnlo~\cite{Frixione:2002ik,Frixione:2010ra} generator, which incorporates the NLO QCD matrix elements into the parton 
shower by interfacing to the~\herwig~\cite{Herwig} program. 
The generator also provides matrix element information which allows a given sample to be reweighted to a different set of anomalous coupling parameters on an event-by-event basis.
The parton
density function (PDF) set CTEQ6.6~\cite{cteq6.6} is used and the
underlying event is modelled with \Jimmy~\cite{Jimmy,ATL-PHYS-PUB-2010-014}. 
\herwig is used to model the hadronization, initial state radiation and QCD final state radiation (FSR). 
\photos~\cite{photos} is used for QED FSR, and \tauola~\cite{tauola} for the $\tau$ lepton decays.

The \WZ production cross section at NLO in $\alpha_s$ as previously defined is calculated 
with the program MCFM~\cite{Campbell:2011} 
to be $\WZtotXsecExpected^{+\WZtotXsecExpectedErrUp}_{\WZtotXsecExpectedErrDw}$ pb.
Electroweak corrections are not considered as they are not relevant at the currently available integrated 
luminosity~\cite{Accomando:2001fn,Accomando:2005xp}.

The background sources for which data-driven methods could not be used were estimated with simulated samples.
The diboson processes $WW$ and $ZZ$ are modelled with \herwig, and $W/Z+\gamma$ with \madgraph~\cite{madgraph} and \pythia~\cite{pythia}.
\mcatnlo~\cite{Frixione:2002ik} is used to model the $t\bar t$ and single top-quark background in the $\W\Z\to e\nu ee$ decay channel.
Whenever LO event generators are used, the cross sections are corrected 
by using k-factors to NLO or NNLO (if available) matrix element calculations~\cite{Frixione:2002ik,Campbell:2011,Anastasiou:2003ds,2002NuPhB,Kauer:2007zz}. 

The response of the ATLAS detector is simulated~\cite{bib:ATLASMCPaper} with \geant4~\cite{geant4}. Small response and efficiency corrections, based on studies in data and simulated control samples, are applied to the simulated samples.
All event samples are simulated with in-time pile-up (multiple $pp$ interactions within a single bunch crossing) and out-of-time pile-up (signals from nearby bunch crossings).
The weights of simulated events are defined such that the distribution of multiple collisions per bunch crossing matches the observation in the data period under consideration.

%% file: ObjectReconstruction.tex
The main physics objects necessary to select \WZ events are electrons, muons, and \met. 
Muons are identified by matching tracks reconstructed in the MS to tracks reconstructed in the ID. Their momenta are calculated by combining information from the two tracks and correcting for energy deposited in the ca\-lo\-ri\-meter. ID tracks that are tagged as muons on the basis of matching with track segments in the MS (`segment-tagged' muons~\cite{ATLAS-CONF-2011-063}) are also included.
Only muons with $\pt >15$~GeV and $|\eta| < 2.5$ are considered. 
Non-prompt muons from hadronic jets are rejected by selecting only isolated muons, requiring the scalar sum of the $\pt$ of tracks within $\Delta R <0.2$  of the muon to be less than 10\% of the muon $\pt$~\cite{ATLAS-CONF-2011-063}.

Electrons are reconstructed by matching clusters found in the electromagnetic calorimeter to tracks in the ID.  Electron candidates must have $E_{\mathrm T} > 15$ GeV, where $E_{\mathrm T}$ is calculated from the cluster energy and track direction. To avoid the transition regions between the calorimeters, the electron cluster must satisfy $|\eta| < 1.37$ or $1.52 < |\eta| < 2.47$. 
Electrons are required to pass the `medium' identification criteria described in Ref.~\cite{EleReco}.
    To ensure isolation,  the sum of the calorimeter energy in a cone of $\Delta R = 0.3$ around the electron candidate, not including the energy of the cluster associated to the candidate itself, must be less than 4 GeV. 

The \met $\mbox{}$ is calculated with reconstructed electrons within $|\eta| < 2.47$, muons within $|\eta |< 2.7$, and jets and calorimeter energy clusters outside of other reconstructed objects within $|\eta| < 4.5$.
The clusters are calibrated as electromagnetic or hadronic energy according to cluster topology.  A small correction avoids double-counting the energy deposited by muons in the calorimeters~\cite{METReco}. 

%% file: EventSelection.tex
At least one single electron or muon trigger is required for the event selection. 
A minimum of one reconstructed vertex, with at least three tracks associated with it, is required to remove non-collision backgrounds. 
The vertex with the largest sum of the $\pT^2$ computed from the associated tracks is selected as 
the primary vertex.
Events with two leptons of the same flavour and opposite charge with an invariant mass within 10 GeV of the $Z$ boson mass are selected. For the $e\nu ee$ and $\mu\nu\mu\mu$ channels more than one lepton pair combination may satisfy this criterion and the pair closest to the \Z\ boson mass is chosen. 
This requirement of a lepton pair consistent with originating from a \Z\ boson reduces much of the background from multijet and top-quark production, and a fraction of the diboson background.

Events are then required to have at least three reconstructed leptons
originating from the primary vertex; 
their longitudinal impact parameters with respect to the primary vertex are required to be less than~{10 mm}. 

The lepton not attributed to the $Z$ boson decay must pass more stringent identification criteria than the leptons attributed to the $Z$ boson, and have $\pt >$ 20~GeV.
Electrons are additionally required to pass the `tight' identification criteria~\cite{EleReco} with cuts 
on the matched track quality, the ratio of the energy measured in the calorimeter to the momentum of the matched track, 
and the detection of transition radiation. Segment-tagged muons may not be used as the third lepton.

Events are required to have $\met>25$ GeV and the transverse mass
of the \W\ boson candidate, $\MT^W$, formed from the $\met$ and the third lepton, 
is required to be greater than 20 GeV. These cuts suppress
the remaining backgrounds from $Z$ and $ZZ$ production. 

At least one of the leptons is
required to have fired the trigger. To ensure that the trigger is well onto the efficiency plateau
above the threshold of the primary single-lepton trigger, trigger-matched leptons are required to have $\pt >$ 20 GeV for muons and 25 GeV for electrons.

%% file: SignalAcceptance.tex
The fiducial efficiency is defined as the ratio of simulated signal events meeting the event selection criteria to the numbers of simulated events\footnote{Contributions from $\tau$ lepton decays are excluded.}  within the defined fiducial phase space region. 
The values for each channel are shown in Table~\ref{table:Acc}. The fraction of selected simulated signal events which come from outside
the fiducial phase space is 13\%. 

\begin{table*}[htbp]
         \centering
        \begin{tabular}{lcccc}
         \hline\hline
          Final State & $eee+\met$  & $ee\mu+\met$ & $e\mu\mu+\met$ & $\mu\mu\mu+\met$ \\
         \hline
         Fiducial Efficiency (\%) & \CWZeeeRel$\pm\CWZeeeRelErr$   & \CWZeemRel$\pm\CWZeemRelErr$    & \CWZemmRel$\pm\CWZemmRelErr$    & \CWZmmmRel$\pm\CWZmmmRelErr$    \\
         \hline\hline
         \end{tabular}
         \caption{Fiducial efficiency per channel. The uncertainty due to simulated sample size and parton distribution functions is shown.}
          \label{table:Acc}
\end{table*}

%% file: Systematics.tex
The total systematic uncertainty on the efficiency is 3--7\% 
depending on the decay channel and is dominated by the uncertainties on the 
electron and muon reconstruction.
These include uncertainties associated with the reconstruction and identification efficiencies, energy scale, 
and isolation. The uncertainties are
determined by comparing simulated events with data in control
regions and are 2--6\% depending on the decay channel. 
The uncertainties on the objects involved in the \met\ calculation are used 
to derive the systematic uncertainties on \met\ following Ref.~\cite{METReco}. Uncertainties in the description of the pile-up conditions by the simulation are also considered.
The total systematic uncertainty on the acceptance of the \met\ and transverse mass cuts due to the imperfect simulation is 1--2\%.

%% file: Background.tex
Data-driven methods are used to estimate the backgrounds from $Z+$jets
and top-quark production. Simulation is used for the remaining background sources,
including $W/Z+\gamma$ events where the photon converts into an electron-positron pair.
The backgrounds from $\WW$ and multijet production are negligible. 
For simulated events, the uncertainties on the theoretical cross section
of the background processes are included in the systematic uncertainty.

In the $\mu\nu ee$, $e\nu\mu\mu$ and $\mu\nu\mu\mu$ channels, the top-quark background contribution 
is evaluated from the average density of events in the side-bands around the $Z$ mass peak after applying all selection cuts except the $Z$ boson mass cut. 
Since the background from top-quark production does not contain a $Z$ boson, this density is used to estimate the background from top-quark production in the signal region within the $Z$ mass window.
The systematic uncertainty is estimated from various cross checks, including a comparison of the difference between the side-band estimate and the prediction within the $Z$ mass window in simulated events. 
This method is not applicable to the $e\nu ee$ channel, 
since the $Z$+jet background dominates the side-bands due to electron misidentification, therefore a simulated event sample is used.

In order to estimate the background from $Z+$jets events, a sample of events containing a $Z$ boson candidate selected as described above and one
``lepton-like'' jet is identified. The lepton-like jet is a lepton candidate which does not explicitly have to satisfy lepton quality ($e$) or
isolation ($\mu$) requirements. 
To ensure that the control sample is as similar to the signal as possible, 
all other event selection criteria, including the $\met$ and $\MT^W$
requirements, are applied.  The background contribution is then estimated by scaling each event 
in the resulting sample by the probability $f(\pt)$ that a ``lepton-like''
jet satisfies the quality or isolation requirements.  
The scaling factor $f(\pt)$ is determined from a data sample of events containing a $Z$ boson plus an extra
lepton-like jet, with a low missing transverse momentum, $\met < 25\GeV$. The validity of extrapolation to high values of $\met$ has been verified with
dijet events from simulation and data. An estimate of the 
systematic uncertainty is derived from the $\met$ extrapolation
in dijet data.

%% file: Results.tex
\begin{table*}[htbp]
  \centering
  \begin{tabular}{lccccc} 
    \hline\hline
     Final State    & $eee+\met$ & $ee\mu+\met$ & $e\mu\mu+\met$ &
    $\mu\mu\mu+\met$   & Combined \\
    \midrule
    Observed        & \WZNSigEventsObservedEEE &
    \WZNSigEventsObservedEEM & \WZNSigEventsObservedEMM &
    \WZNSigEventsObservedMMM  & \WZNSigEventsObserved \\
    \midrule
    $ZZ$  
    & \WZNZZBkgEventsExpectedEEE$\pm$\WZNZZBkgEventsExpectedErrStatEEE
    & \WZNZZBkgEventsExpectedEEM$\pm$\WZNZZBkgEventsExpectedErrStatEEM   
    & \WZNZZBkgEventsExpectedEMM$\pm$\WZNZZBkgEventsExpectedErrStatEMM 
    &
    \WZNZZBkgEventsExpectedMMM$\pm$\WZNZZBkgEventsExpectedErrStatMMM
  & \WZNZZBkgEventsExpected$\pm$\WZNZZBkgEventsExpectedErrStat$\pm$\WZNZZBkgEventsExpectedErrSys \\
    $W/Z+$jets     
    & \WZNZjetsDDBkgEventsExpectedEEE$\pm$\WZNZjetsDDBkgEventsExpectedErrStatEEE
    & \WZNZjetsDDBkgEventsExpectedEEM$\pm$\WZNZjetsDDBkgEventsExpectedErrStatEEM   
    & \WZNZjetsDDBkgEventsExpectedEMM$\pm$\WZNZjetsDDBkgEventsExpectedErrStatEMM 
    &
    \WZNZjetsDDBkgEventsExpectedMMM$\pm$\WZNZjetsDDBkgEventsExpectedErrStatMMM
 & \WZNZjetsDDBkgEventsExpected$\pm$\WZNZjetsDDBkgEventsExpectedErrStat$^{+\WZNZjetsDDBkgEventsExpectedErrSysUp}_{-\WZNZjetsDDBkgEventsExpectedErrSysDown}$ \\
    Top           
    & \WZNTopBkgEventsExpectedEEE$\pm$\WZNTopBkgEventsExpectedErrStatEEE
    & \WZNTopBkgEventsExpectedEEM$\pm$\WZNTopBkgEventsExpectedErrStatEEM   
    & \WZNTopBkgEventsExpectedEMM$\pm$\WZNTopBkgEventsExpectedErrStatEMM 
    &
    \WZNTopBkgEventsExpectedMMM$\pm$\WZNTopBkgEventsExpectedErrStatMMM
 & \WZNTopBkgEventsExpected$\pm$\WZNTopBkgEventsExpectedErrStat$\pm$\WZNTopBkgEventsExpectedErrSys \\   
    $W/Z+\gamma$  
    & \WZNZgammaBkgEventsExpectedEEE$\pm$\WZNZgammaBkgEventsExpectedErrStatEEE
    & --   
    & \WZNZgammaBkgEventsExpectedEMM$\pm$\WZNZgammaBkgEventsExpectedErrStatEMM
    & -- 
    & \WZNZgammaBkgEventsExpected$\pm$\WZNZgammaBkgEventsExpectedErrStat$\pm$\WZNZgammaBkgEventsExpectedErrSys \\ 
    \midrule
    Total Background        & \WZNBkgEventsExpectedEEE$\pm$\WZNBkgEventsExpectedErrStatEEE
    & \WZNBkgEventsExpectedEEM$\pm$\WZNBkgEventsExpectedErrStatEEM
    & \WZNBkgEventsExpectedEMM$\pm$\WZNBkgEventsExpectedErrStatEMM
    &
    \WZNBkgEventsExpectedMMM$\pm$\WZNBkgEventsExpectedErrStatMMM
    & \WZNBkgEventsExpected$\pm$\WZNBkgEventsExpectedErrStat$^{+\WZNBkgEventsExpectedErrUpSys}_{-\WZNBkgEventsExpectedErrDwSys}$ \\
    Expected Signal        
    & \WZNSigEventsExpectedEEE$\pm$\WZNSigEventsExpectedErrStatEEE    
    & \WZNSigEventsExpectedEEM$\pm$\WZNSigEventsExpectedErrStatEEM 
    & \WZNSigEventsExpectedEMM$\pm$\WZNSigEventsExpectedErrStatEMM 
    &
    \WZNSigEventsExpectedMMM$\pm$\WZNSigEventsExpectedErrStatMMM
    & \WZNSigEventsExpected$\pm$\WZNSigEventsExpectedErrStat$\pm$\WZNSigEventsExpectedErrSys \\
    Total Expected Events        
    & \WZNSumSigEventsExpectedEEE$\pm$\WZNSumSigEventsExpectedErrStatEEE    
    & \WZNSumSigEventsExpectedEEM$\pm$\WZNSumSigEventsExpectedErrStatEEM 
    & \WZNSumSigEventsExpectedEMM$\pm$\WZNSumSigEventsExpectedErrStatEMM 
    &
    \WZNSumSigEventsExpectedMMM$\pm$\WZNSumSigEventsExpectedErrStatMMM
    & \WZSumNSigEventsExpected$\pm$\WZNSumSigEventsExpectedErrStat$^{+\WZNSigEventsExpectedErrSysUp}_{-\WZNSigEventsExpectedErrSysDw}$ \\
    \hline\hline
  \end{tabular}
  \caption{Summary of observed events and expected signal and background
contributions for the four trilepton channels and their combination.
 Statistical uncertainties are shown for the individual channels, and
both statistical and systematic uncertainties are shown for the
combined channel. Expected signal (\WZ) and background events from $ZZ$ and $W/Z+\gamma$ are predicted from MC simulation. Data-driven background estimation methods are
used for $W/Z+$jets for all decay channels. For backgrounds with top-quark decays, data-driven estimates are used for the $\mu\mu\mu$, $e\mu\mu$ and $ee\mu$ channels whereas MC simulation is used for the $eee$ channel. $W/Z+\gamma$ does not contribute to the $ee\mu$ and $\mu\mu\mu$ channels.
}
  \label{ta:selected_data_MC}
\end{table*}

 \begin{figure}
 \includegraphics[width=0.5\textwidth]{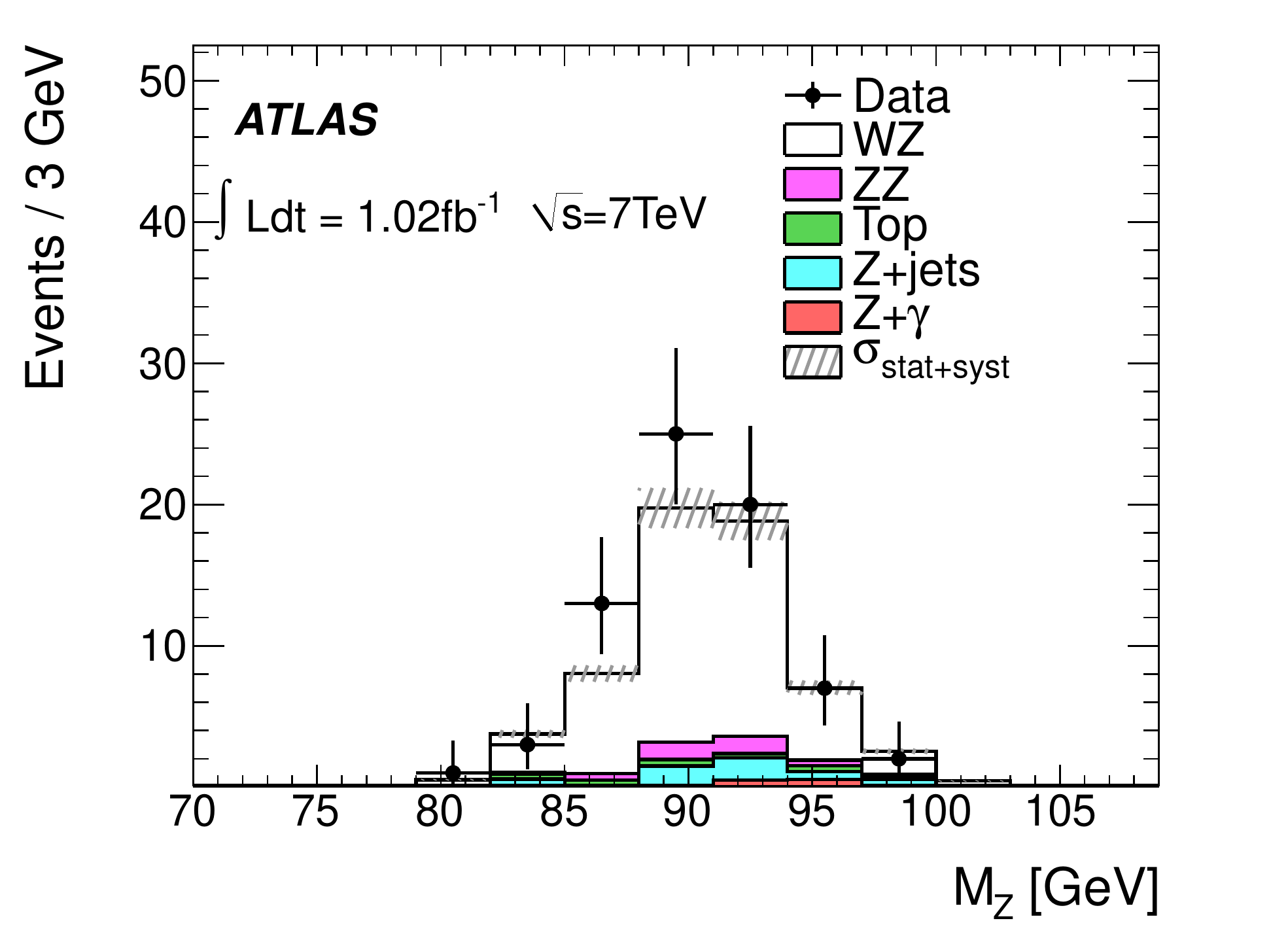}
 \caption{\small The invariant mass of the lepton pair attributed to the $Z$ boson in candidate events after the full selection.
The stacked histograms represent the predictions from simulation or, where applicable, data-driven estimates including the statistical and systematic uncertainty shown by shaded bands.
The shape of the top-quark background is taken from simulation.
   }
 \label{fig:wz-cand}
 \end{figure} 

\begin{figure}
 \includegraphics[width=0.5\textwidth]{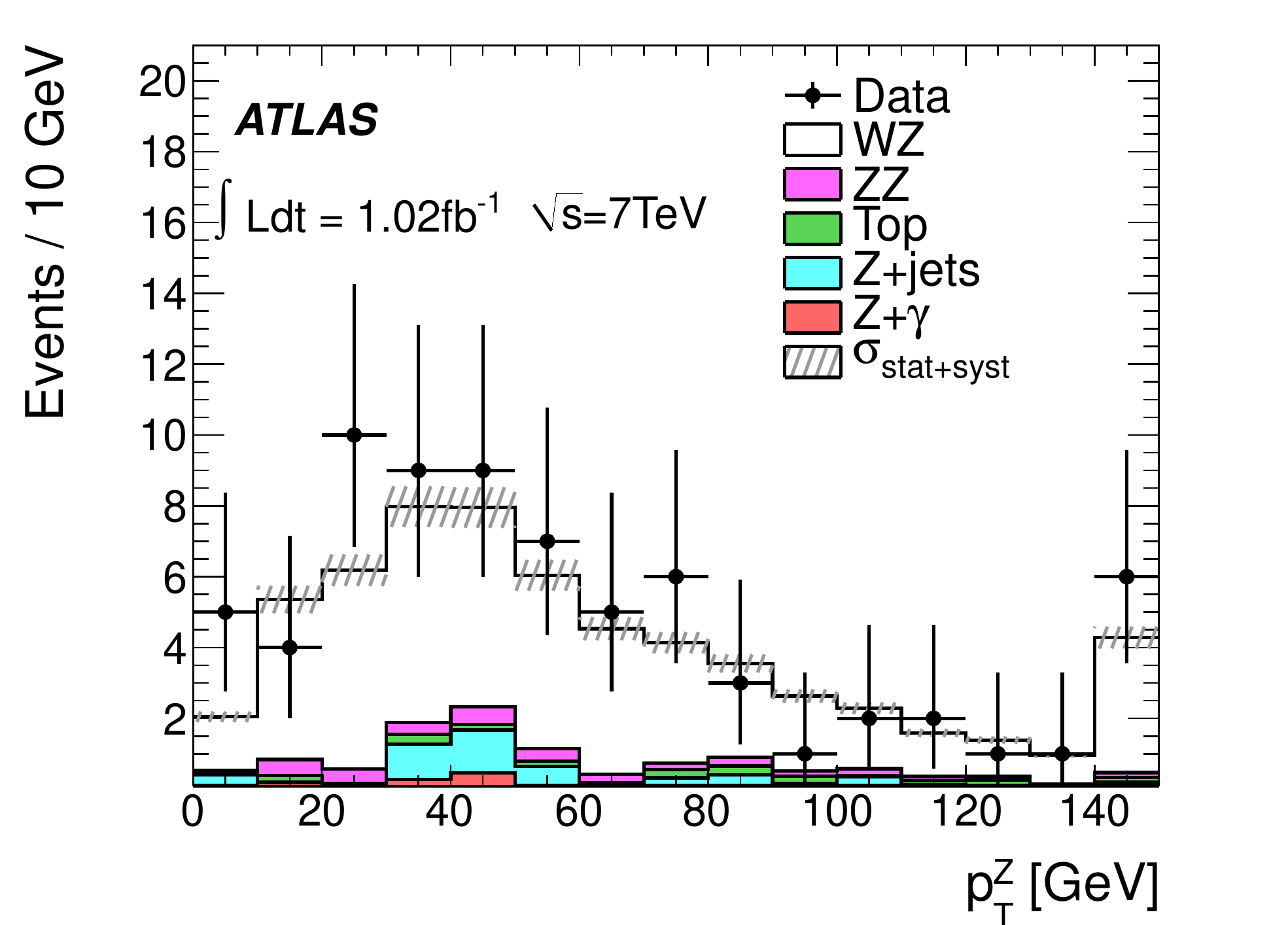}
 \caption{\small The transverse momentum of $Z$ bosons in candidate events after full selection.
The stacked histograms represent the predictions from simulation or, where applicable, data-driven estimates including the statistical and systematic uncertainty shown by shaded bands. The last bin includes the overflow. The shape of the top-quark background is taken from simulation.
   }
 \label{fig:wz-cand2}
 \end{figure}

The numbers of expected and observed events after the full selection are shown in Table~\ref{ta:selected_data_MC}.
A total of \WZNSigEventsObserved\ \WZ\ candidates are observed in data, with 
$\WZNBkgEventsExpected\pm\WZNBkgEventsExpectedErrStat\mathrm{(stat.)}^{+\WZNBkgEventsExpectedErrUpSys}_{-\WZNBkgEventsExpectedErrDwSys}\mathrm{(syst.)}$ 
expected background events. The expected signal events shown in the table include the contribution from $\tau$ lepton decays 
into electrons or muons. 
The discrepancy between channels in the number of observed to expected
events is consistent with a statistical fluctuation at the 16\% level.
The invariant mass and the transverse momentum of the $Z$ boson in \WZ
candidate events are shown in Figures~\ref{fig:wz-cand}
and~\ref{fig:wz-cand2}, respectively.

The fiducial cross section is calculated from
\begin{equation}\label{eq:xsecfid}
\sigma_{W\!Z\to\ell\nu\ell\ell}^\mathrm{fid} =
\frac{ N^\mathrm{obs}_{\ell\nu\ell\ell} - N^\mathrm{bkg}_{\ell\nu\ell\ell}}{\mathcal{L} \times C_{W\!Z\to\ell\nu\ell\ell}}\times\left(1-\frac{N^\mathrm{MC}_{\tau}}{N^\mathrm{MC}_\mathrm{sig}}\right)
\end{equation}
\noindent where $N^\mathrm{obs}_{\ell\nu\ell\ell}$ and $N^\mathrm{bkg}_{ \ell\nu\ell\ell}$ are the
numbers of observed and background events, $\mathcal{L}$ the integrated
luminosity and \linebreak $C_{W\!Z\to\ell\nu\ell\ell}$ is the fiducial efficiency 
defined above. The last term corrects for the $\tau$ lepton contribution estimated from the selected simulated
signal sample, where
$N^\mathrm{MC}_{\tau}$ is the number of \WZ\ events with at least one of the bosons decaying to a $\tau$ lepton and
$N^\mathrm{MC}_{\mathrm{sig}}$ is the number of \WZ\ events with decays into any lepton flavour. For each final state, the simulated signal samples include $W$ and $Z$ bosons decaying into $\tau$ leptons. The contribution from $\tau$ lepton decays is 3.7\% summing over all channels.

The total cross section is calculated as 
\begin{equation}\label{eq:xsectot}
\sigma_{W\!Z}^\mathrm{tot} =
\frac{\sigma_{W\!Z\to\ell\nu\ell\ell}^\mathrm{fid}}
     {\mathcal{B}(W\!Z\to\ell\nu\ell\ell) \times A_{W\!Z\to\ell\nu\ell\ell}}
\end{equation}
where $\mathcal{B}(W\!Z\to\ell\nu\ell\ell)$ is the branching ratio for a \W\ boson to decay to
$\ell\nu$ and a $Z$ boson to decay to $\ell\ell$, and 
$A_{WZ\to\ell\nu\ell\ell}$ is the ratio of the number of events
within the fiducial phase space region to the number of events
within $66\GeV <m_{\ell\ell}<116\GeV$. 
This ratio $A_{W\!Z\to\ell\nu\ell\ell}$ is calculated at NLO to be $\AWZ\pm\AWZErrorComb$ 
using \mcfm~\cite{Campbell:2011} with PDF set CTEQ6.6, where 
the uncertainty arises from the statistical error due to the sample size 
in the \mcfm\ integration (\AWZrelErrorStat\%) and 
parton distribution function uncertainty (\AWZrelErrorComb\%).

The cross section is determined by minimizing a negative log-likelihood function to
combine the four channels. Systematic uncertainties are included as
Gaussian-constrained nuisance parameters. 
For each systematic uncertainty, correlations between signal and background predictions are taken into account. All uncertainties are allowed to vary simultaneously in the fit.

The measurements of the combined fiducial cross section for the \WZ bosons
    decaying directly into electrons and muons, and the total inclusive cross
    section, are
\begin{eqnarray}\label{eq:fid}\hskip -.4cm
\sigma_{W\!Z\to\ell\nu\ell\ell}^\mathrm{fid} &=&
\WZfidXsec^{+\WZfidXsecStatErrUp}_{-\WZfidXsecStatErrDw}\mathrm{(stat.)}
^{+\WZfidXsecSysErrUp}_{-\WZfidXsecSysErrDw}\mathrm{(syst.)}
^{+\WZfidXsecLumiErrUp}_{-\WZfidXsecLumiErrDw}\mathrm{(lumi.)}\,\mathrm{fb}
\end{eqnarray}
\begin{eqnarray}\label{eq:tot}\hskip -.4cm
\sigma_{W\!Z}^\mathrm{tot} &=&
\WZtotXsec^{+\WZtotXsecStatErrUp}_{-\WZtotXsecStatErrDw}\mathrm{(stat.)}
^{+\WZtotXsecSysErrUp}_{-\WZtotXsecSysErrDw}\mathrm{(syst.)}
^{+\WZtotXsecLumiErrUp}_{-\WZtotXsecLumiErrDw}\mathrm{(lumi.)}\,\mathrm{pb}.
\end{eqnarray}
The latter can be compared with the SM expectation, $\WZtotXsecExpected^{+\WZtotXsecExpectedErrUp}_{\WZtotXsecExpectedErrDw}$ pb, calculated with MCFM~\cite{Campbell:2011}.

\input{ResultsSubSecTGC}

\section{Conclusion} \label{subsec:Conclusion}
A measurement of the \WZ production cross section has been performed
   using final states with electrons and muons, in LHC $pp$ collisions at
   $\sqrt{s}=7$ TeV with ATLAS. In data with an integrated luminosity of \lumi~\ifb, a total of \WZNSigEventsObserved\ candidates is observed with a background
expectation of
$\WZNBkgEventsExpected\pm\WZNBkgEventsExpectedErrStat\mathrm{(stat.)}^{+\WZNBkgEventsExpectedErrUpSys}_{-\WZNBkgEventsExpectedErrDwSys}\mathrm{(syst.)}$ events. The SM expectation for the number of signal events is $\WZNSigEventsExpected\pm\WZNSigEventsExpectedErrStat\mathrm{(stat.)}\pm\WZNSigEventsExpectedErrSys\mathrm{(syst.)}$. The fiducial and total cross sections determined in the present work are given in equations~\ref{eq:fid} and~\ref{eq:tot}, respectively.
The total cross section is in good agreement with the SM expectation. 
Limits on the anomalous triple gauge couplings $\Delta g^{Z}_{1}$, $\Delta\kappa_Z$ and $\lambda_Z$ are reported and 
the results are consistent with zero, as expected from the SM.

%% file: ResultsSubSecTGC.tex
In order to set limits on the anomalous coupling parameters, a frequentist approach~\cite{frequentist} is used with the profile likelihood ratio used as the test statistic.
The limits are set separately on each parameter with the other couplings fixed to their SM values. 
A reweighting procedure is used to predict the numbers of expected events as functions of the parameter being studied.
The uncertainties on the signal acceptance and efficiency and on the background estimates are included as nuisance parameters with Gaussian constraints in the likelihood function.
The 95\% confidence interval (C.I.) is defined as the range(s) of the coupling parameter(s) for which at least 5\% of randomly generated pseudo-experiments result in a smaller value of the profile likelihood ratio than is observed with the data.

\begin{figure}[htbp]
 \begin{center}
 \includegraphics[width=.5\textwidth]{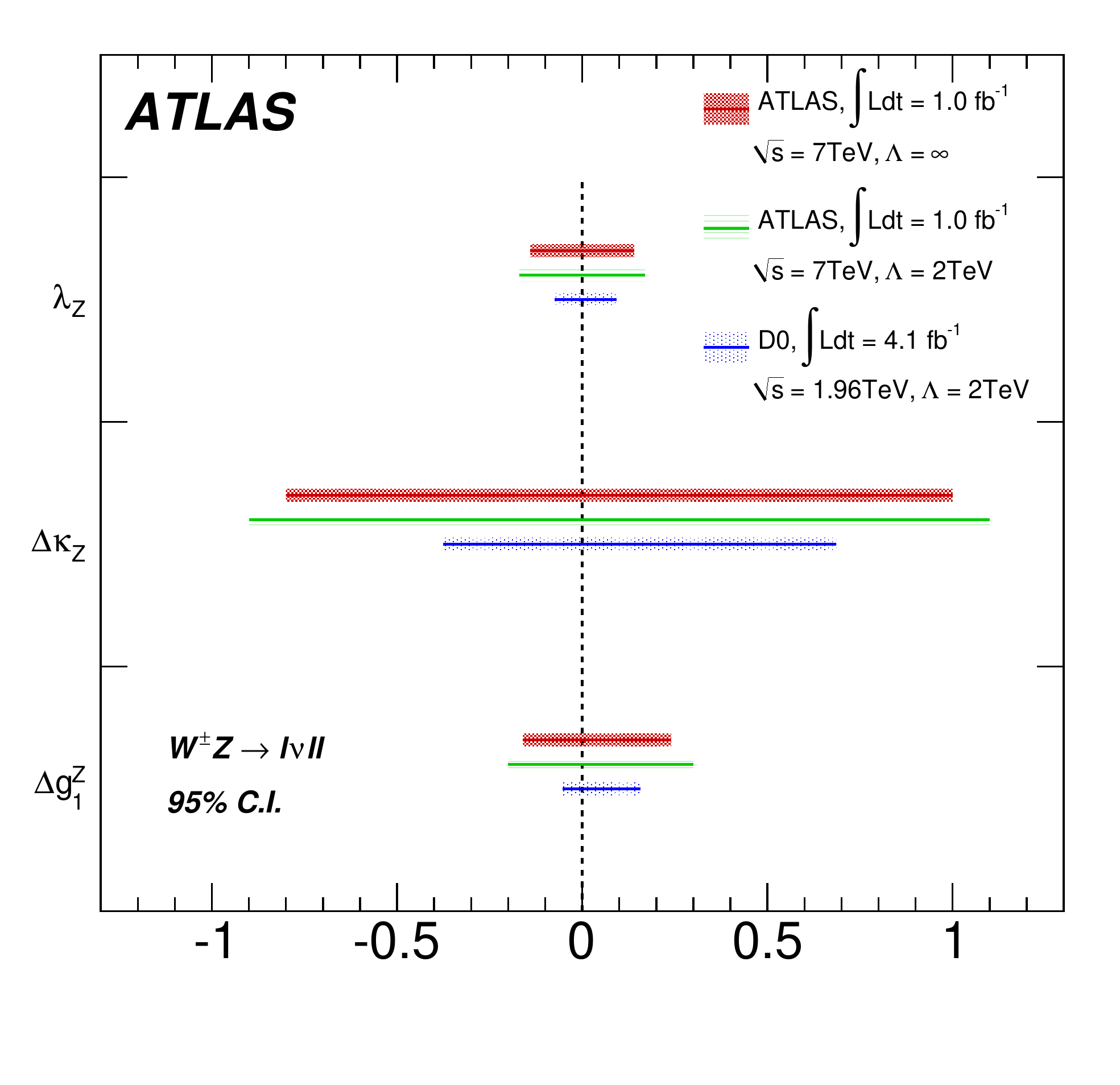}
\caption{\small 95\% C.I. for anomalous couplings from ATLAS and D\O\ experiments. ATLAS limits are extracted from a fit to the cross section while the D\O~\cite{D0limits} limits are extracted from a fit to the $\pT(Z)$ spectrum. Luminosities, centre-of-mass energy and cut-off $\Lambda$ are shown.}
 \label{fig:combined}
 \end{center}
 \end{figure}

The observed and expected 95\% C.I. for the anomalous couplings are summarized in Table \ref{table:lim}.
The observed limits are compared with D\O\ results from \WZ production in Figure~\ref{fig:combined}.
Other results on anomalous couplings from $W^+W^-$ production can be found in Refs.~\cite{bib:OPALWW,bib:DELPHIWW,bib:ALEPHWW,bib:L3WW, CMSlimitsWW, D0limitsWW, CDFlimitsWW}.
Significant improvements in these limits are expected with more integrated luminosity and refined extraction methods which take advantage of the differential spectra of kinematic quantities. 
The anomalous couplings influence the kinematic properties of \WZ events and thus the fiducial efficiency. The $C_{W\!Z}$ variation within the measured aTGC limits results maximally in a 3$\%$ decrease of the fiducial cross section.

\begin{table}[htbp]
\renewcommand{\arraystretch}{1.4}
	 \centering
	 \begin{tabular}{lccc}
	 \hline	 \hline
	 Coupling & Observed & Observed & Expected \\
	 & ($\Lambda=2$ TeV) & ($\Lambda=\infty$) & ($\Lambda=\infty$)\\
	 \hline
	 $\Delta g_1^Z$ &
	 $\left[\obsglowtwotev,  \obsguptwotev\right]$ &
	 $\left[\obsglow,  \obsgup\right]$ &
	 $\left[\expglow, \expgup \right]$\\

	 $\Delta \kappa_Z$ & 
	 $\left[\obskappalowtwotev,  \obskappauptwotev\right]$ &
	 $\left[\obskappalow,  \obskappaup\right]$ &
	 $\left[\expkappalow, \expkappaup \right]$\\

	 $\lambda_{Z}$ &  
	 $\left[\obslambdalowtwotev,  \obslambdauptwotev\right]$ &
	  $\left[\obslambdalow,  \obslambdaup\right]$ &
	 $\left[\explambdalow, \explambdaup \right]$\\

	 \hline	 \hline
	 \end{tabular}
	    \caption{Observed and expected 95\% C.I. for the anomalous couplings $\Delta g_1^Z$, $\Delta \kappa_Z$, and $\lambda_Z$.
	    Expected experimental limits assume SM values. 
            }
          \label{table:lim}
\end{table}

%% file: atlas_authlist.tex
\begin{flushleft}
{\Large The ATLAS Collaboration}

\bigskip

G.~Aad$^{\rm 48}$,
B.~Abbott$^{\rm 111}$,
J.~Abdallah$^{\rm 11}$,
A.A.~Abdelalim$^{\rm 49}$,
A.~Abdesselam$^{\rm 118}$,
O.~Abdinov$^{\rm 10}$,
B.~Abi$^{\rm 112}$,
M.~Abolins$^{\rm 88}$,
H.~Abramowicz$^{\rm 153}$,
H.~Abreu$^{\rm 115}$
% ...
\bigskip

% ...
$^{10}$ Institute of Physics, Azerbaijan Academy of Sciences, Baku, Azerbaijan\\
$^{11}$ Institut de F\'isica d'Altes Energies and Departament de F\'isica de la Universitat Aut\`onoma  de Barcelona and ICREA, Barcelona, Spain\\
% ...
$^{48}$ Fakult\"{a}t f\"{u}r Mathematik und Physik, Albert-Ludwigs-Universit\"{a}t, Freiburg i.Br., Germany\\
$^{49}$ Section de Physique, Universit\'e de Gen\`eve, Geneva, Switzerland\\
% ...
$^{88}$ Department of Physics and Astronomy, Michigan State University, East Lansing MI, United States of America\\
% ...
$^{111}$ Homer L. Dodge Department of Physics and Astronomy, University of Oklahoma, Norman OK, United States of America\\
$^{112}$ Department of Physics, Oklahoma State University, Stillwater OK, United States of America\\
% ...
$^{115}$ LAL, Univ. Paris-Sud and CNRS/IN2P3, Orsay, France\\
% ...
$^{118}$ Department of Physics, Oxford University, Oxford, United Kingdom\\
% ...
$^{153}$ Raymond and Beverly Sackler School of Physics and Astronomy, Tel Aviv University, Tel Aviv, Israel\\
% ...
\end{flushleft}
